\documentclass[a4paper,12pt]{article}

\pdfoutput=1
\usepackage{jheppub}

\title{Note On The Dilaton Effective Action And Entanglement Entropy}

\author[]{Shamik Banerjee}
\affiliation[]{Kavli Institute for the Physics and Mathematics of the Universe (WPI), Todai Institutes for Advanced Study, The University of Tokyo, Kashiwa, Chiba 277-8583, Japan}

\emailAdd{banerjeeshamik.phy@gmail.com}

\abstract{In this note we do the analysis of entanglement entropy more carefully when the non-conformal theory flows to a non-trivial IR fixed point. In particular we emphasize the role of the trace of the energy-momentum tensor in these calculations. We also compare the current technique for evaluating the entanglement entropy, particularly the Green's function method for gaussian theories, with the dilaton effective action approach and show that they compute identical quantities. As a result of this, the dilaton effective action approach can be thought of as an extension of Green's function technique to interacting theories.  }

\begin{document}

%\preprint{23}
\begin{flushright} 
\small{IPMU14-0137} 
\end{flushright}

\maketitle
\flushbottom

\section{Two Dimensional Non-Conformal Theory}
In \cite{Banerjee:2014daa} we showed that the dilation effective action on the cone can be used to compute the entanglement entropy of a non-conformal field theory. Let us review the argument now. Away from criticality the theory is no longer conformally invariant. The conformal invariance can be restored by coupling the theory to a background dilation field \cite{Komargodski:2011vj, Komargodski:2011xv}.\footnote{See also \cite{Luty:2012ww,Elvang:2012st,Schwimmer:2010za,Nakayama:2011wq}} Now the dilation couples to the trace of the energy momentum tensor of the theory. So one can compute the trace of the energy momentum from the dilation effective action. In particular if the theory is coupled to a constant dilation background field then the effective action gives the integrated trace of the energy momentum tensor. The integrated trace of the energy-momentum measures the response of the theory to a constant scale transformation. This fact is useful for computing entanglement entropy, see for example \cite{Holzhey:1994we,Calabrese:2004eu,Ryu:2006ef,Myers:2010tj}. So let us consider a two dimensional non-conformal theory which was treated in \cite{Calabrese:2004eu,Calabrese:2009qy}. The theory flows from a UV fixed point with central charge $c_{UV}$ to an IR fixed point with central charge $c_{IR}$. Entanglement entropy, $S_{EE}$, can be computed by the replica trick and is given by, 
\begin{equation}\label{ent}
S_{EE} =  n\frac{\partial}{\partial n} (F(n) - nF(1)) \ |_{n=1}
\end{equation}
where $F(n) = -ln Z(n)$, is the free energy of the Euclidean field theory on a space with conical singularities. The angular excess at each conical singularity is given by $2\pi(n-1)$. The detailed geometry of the singular space depends on the choice of the subsystem for which we are computing the entanglement entropy and the geometry of the background space-time. Let us choose the subsystem to be the infinite half-line and the background space-time to be the two dimensional Euclidean plane. In that case the geometry of the replica space is just that of a flat cone with one conical singularity and angular excess given by $2\pi(n-1)$. In such a space, the trace of the energy momentum of the non-conformal theory was calculated in \cite{Calabrese:2004eu,Calabrese:2009qy} and is given by, 
\begin{equation}\label{CC1}
\int_{cone} (<T^{\mu}_{\mu}>_n - <T^{\mu}_{\mu}>_1) =  -\pi n \frac{c_{UV} - c_{IR}}{6} (1-\frac{1}{n^2})
\end{equation}
where $T^{\mu}_{\mu}$ is the trace of the energy-momentum. $<T_{\mu}^{\mu}>_{n}$ denotes the expectation value of the trace on the cone and $<T^{\mu}_{\mu}>_{1}$ denotes the expectation value of the trace on the plane. The above formula computes the contribution of the conical singularity to the trace of the energy-momentum of the non-conformal theory. Let us first show that this result can also be obtained by coupling the theory to a constant background dilation field on the cone. 

The effective action for a constant background dilation field $\tau$ is given by, \footnote{The reader should note that this is the universal part of the dilaton effective action which is uniquely fixed by anomaly matching. In the case of an infinite half-line the non-universal or Weyl-invariant part of the dilaton effective action does not contribute because there is no other length scale in the problem. If the subsystem is a finite interval then the non-universal part has to be calculated on the replica geometry and it will give finite size corrections to the entropy. Some of these terms were considered in \cite{Banerjee:2014daa} for higher dimensional theories and were shown to give rise to interesting universal terms. They will not be discussed in this note.}
\begin{equation}\label{ea1}
F(n,\tau) = -\frac{c_{UV} - c_{IR}}{24\pi} \  \frac{1}{2}(1+ \frac{1}{n}) \ \tau  \int_{cone} \sqrt h R(h)
\end{equation}
where the integral is done on the flat cone as described before. In order to arrive at this form of the effective action we have used the trace anomaly matching \cite{Komargodski:2011vj, Komargodski:2011xv} condition and also used the fact that the trace of the energy-momentum for a conformal field theory of central charge $c$ on the cone is given by \cite{Holzhey:1994we, Calabrese:2004eu},
\begin{equation}\label{cone1}
\int_{cone} \sqrt h <T^{\mu}_{\mu}> = \frac{c}{24\pi} \ \frac{1}{2}(1+ \frac{1}{n})\int_{cone} \sqrt h R(h)
\end{equation}
Now the trace of the energy momentum tensor in the non-conformal theory is the coefficient of the term linear in $\tau$.\footnote{Please see section-$5$ of \cite{Banerjee:2014daa} for a discussion and derivation of this. } Using this fact, one can reproduce Eqn-\ref{CC1} by noting that on the cone,
\begin{equation}
\int_{cone} \sqrt{h} R(h) = 4\pi (1-n)
\end{equation}
This gives us exactly the answer of \cite{Calabrese:2004eu} for the trace of the energy-momentum tensor of a non-conformal theory on the cone (the reader should note that our convention for the energy-momentum tensor differs from that of \cite{Calabrese:2004eu} by an overall minus sign and a factor of $2\pi$). 

Now let ${\mu}$ denote the mass scale associated with the relevant operator which causes the RG flow. Since $\mu$ is the only dimensionful parameter associated with the theory a scale transformation is equivalent to a change in the parameter $\mu$. So one can write,
\begin{equation}
\mu \frac{d}{d\mu} S_{EE} = n\frac{\partial}{\partial n} |_{n=1} \ (\mu\frac{d}{d\mu} F(n) - n \mu\frac{d}{d\mu}F(1))
\end{equation}
where
\begin{equation}
\mu\frac{d}{d\mu} F = -\int \sqrt{h} <T^{\mu}_{\mu}>
\end{equation}
Now using the result for the trace of the energy-momentum we arrive at the result,
\begin{equation}\label{cc}
\mu\frac{d}{d\mu} S_{EE} = - \frac{c_{UV} - c_{IR}}{6} 
\end{equation}
This is the same result as was obtained in \cite{Calabrese:2004eu} by different method. 

Now from this result one can reconstruct the result of the entanglement entropy of an infinite half-line. The result can be written as,
\begin{equation}\label{J1}
S_{EE} = -\frac{c_{UV}}{6} ln (\mu a) + \frac{c_{IR}}{6} ln(\mu L_{IR})
\end{equation}
where $a$ is the short-distance cutoff and $L_{IR}$ is an infrared cutoff. The IR cutoff appears because the entanglement entropy of the IR-CFT is divergent for an infinite half-line. 

When the IR theory is trivial, $c_{IR} = 0$, the second term in $(\ref{J1})$ is not there. This gives us back the answer of \cite{Calabrese:2004eu}, which was also obtained in \cite{Banerjee:2014daa} by using the dilaton background field. The answer, when there is a non-trivial IR-CFT, was proposed by John Cardy in a private correspondence. This answer does not appear in print, but it is implicit in the discussion of \cite{Calabrese:2004eu}.\footnote{See also the footnote-$30$ of \cite{Calabrese:2004eu}.}

\subsection{Relation with the current technique of calculating entanglement entropy}
Let us consider a massive scalar filed of mass $m$ in two dimensions described by the Euclidean action, 
\begin{equation}
S = \frac{1}{2}\int((\partial\phi)^2 + m^2\phi^2)
\end{equation}
We want to compute the entanglement entropy of a subsystem which want to keep arbitrary. It could be an infinite half-line or it could be an interval of finite length. In order to do this one has to compute the free energy of this theory on a space with conical singularities. One way to do this is to use the identity \cite{Calabrese:2004eu,Casini:2009sr},
\begin{equation}\label{c1}
\frac{\partial}{\partial m^2} ln Z_n = - \frac{1}{2} \int G_n(\vec r, \vec r) d^2 \vec r
\end{equation}
where $Z_n$ is the Euclidean partition function of the theory on a space with conical singularities and the angular excess at each singularity has value $2\pi(n-1)$.  $G_n(\vec r, \vec r \ ')$ is the Green's function of the operator $(-\nabla^2 + m^2)$, on the singular space. The trick is to compute the Green's function on the conical geometry and compute the free energy using that. Now instead of doing this one could also use the following identity,
\begin{equation}\label{c2}
m^2 \frac{\partial}{\partial m^2} ln Z_n = -\frac{1}{2} \frac{\partial}{\partial\tau} |_{\tau = 0} \ ln Z_n(\tau)
\end{equation}
where, $-lnZ_n(\tau)$, is the free energy computed on the cone for the theory defined by the euclidean action,
\begin{equation}\label{c3}
S(\tau) = \frac{1}{2}\int((\partial\phi)^2 + m^2 e^{-2\tau}\phi^2)
\end{equation}
Now this is precisely the coupling of the dilaton to the massive theory. So we can interpret the number $\tau$ as a constant background dilaton field. This shows that we can calculate the entanglement entropy once we know the dilaton effective action on the cone. 

Now $(\ref{c1})$ is valid only for gaussian theories but $(\ref{c2})$ can be easily generalised to any theory. The L.H.S of $(\ref{c2})$ computes the trace of the energy-momentum of the massive theory. We have seen that this is enough to compute the entanglement entropy \cite{Calabrese:2009qy}. For a general field theory the dilaton is coupled as \cite{Komargodski:2011xv},
\begin{equation}\label{d1}
S(\tau) = S_{UV-CFT} + \int g(\Lambda e^{\tau}) \Lambda^{2-\Delta} O_{\Delta}(x)
\end{equation}
where $g(\Lambda)$ is the dimensionless coupling constant and $O_{\Delta}$ is the relevant operator of dimension $\Delta$ which causes the RG flow. $\Lambda$ is the scale at which the theory is defined. Although we have written this for two dimensional theories the same form of the coupling holds in higher dimensions with $2$ replaced by the number of space-time dimensions. So if we can calculate the effective action on the cone for the theory defined by $(\ref{d1})$ then we will be able to calculate the entanglement entropy. Precisely this thing was done in the previous section for an infinite half-line. Hence we can think of the dilaton effective action approach as an extension of the Green's function method to interacting theories.

Dilaton effective action has two parts. One is the universal part which is completely determined by the trace anomaly matching \cite{Komargodski:2011vj, Komargodski:2011xv} and the other part is the non-universal or Weyl invariant part. The non-universal part is not determined by symmetry and one needs to calculate this by some other method. In the previous section we have calculated the entanglement entropy when the sub system is an infinite half-line. If the system is instead a finite interval then there will be finite size corrections to this \cite{Calabrese:2004eu,Casini:2009sr} and the non-universal or Weyl invariant part of the dilaton effective action captures this part. Since the non-universal part of the dilaton effective action depends on the details of the RG flow or the system under consideration, it is unlikely that there will be a general closed form answer for the finite size corrections which will be valid for any system. This is supported by the known results \cite{Calabrese:2004eu,Casini:2009sr} for Gaussian theories. 

\section{Higher Dimensional Field Theories}
The method of dilaton field can be applied to higher dimensional field theories and the logic is essentially the same as in the two dimensional case. We can consider a UV-CFT deformed by a relevant operator and let $\mu$ denote the mass scale associated with the operator. Now a scale transformation is equivalent to a change in the parameter $\mu$ and we can write,
\begin{equation}\label{4d1}
\mu \frac{d}{d\mu} S_{EE} = n\frac{\partial}{\partial n} |_{n=1} \ (\mu\frac{d}{d\mu} F(n) - n \mu\frac{d}{d\mu}F(1))
\end{equation}
where
\begin{equation}\label{4d2}
\mu\frac{d}{d\mu} F = -\int \sqrt{h} <T^{\mu}_{\mu}>
\end{equation}
We can now calculate the integrated trace of the energy-momentum from the dilaton effective action. In four dimensions, the universal part of the effective action for a constant background dilaton field is given by, \footnote{As was discussed in section-$6$ of \cite{Banerjee:2014daa}, this expression gives us the correct answer only when the angular excess is infinitesimally small. For general $n$ the central charges should be replaced by their effective values which depend on $n$. Unfortunately this is not known for higher dimensional field theories and so we can only calculate the entanglement entropy in this way but not the Renyi entropy.}
\begin{equation}
F_{universal}(\tau) = - \tau \int d^4 x \sqrt h \ (\frac{c_{UV}-c_{IR}}{16\pi^2} \ W^2 - 2 (a_{UV}-a_{IR}) E_4)
\end{equation}
where $W^2$ and $E_4$ are the Weyl tensor squared and the four dimensional Euler density defined as,
\begin{equation}
W^2 = R_{abcd}R^{abcd} - 2R_{ab}R^{ab} + \frac{1}{3} R^2
\end{equation}
\begin{equation}
E_4 = \frac{1}{32\pi^2} (R_{abcd}R^{abcd} - 4R_{ab}R^{ab} + R^2)
\end{equation}
$(a_{UV},c_{UV})$ and $(a_{IR}, c_{IR})$ are the central charges of the UV and IR-CFTs, respectively. 

One can read off the universal part of the trace of the energy momentum from the universal part of the dilaton effective action and is given by,
\begin{equation}\label{4d3}
\int d^4 x \sqrt h <T^{\mu}_{\mu}>_{universal} \ = - \int d^4 x \sqrt h \ (\frac{c_{UV}-c_{IR}}{16\pi^2} \ W^2 - 2 (a_{UV}-a_{IR}) E_4)
\end{equation}
The trace has many more terms which come from the Weyl invariant part of the dilaton effective action and those have to be computed in some other way. The effect of a class of these terms were considered in \cite{Banerjee:2014daa}. 

Let us now calculate the entanglement entropy coming from the universal part of the trace. We consider a four dimensional space-time whose spatial slice is a cylinder, $S^2\times R^1$, of radius $R$. So the Euclideanized geometry is simply $R^2\times S^2$ and we choose the subsystem to be half of the cylinder, $R^{+}\times S^2$. So the replica geometry is just a two dimensional flat cone with only one conical singularity times the two sphere $S^2$. We can compute the entanglement entropy from $(\ref{4d1})$, $(\ref{4d2})$, $(\ref{4d3})$ and we arrive at,
\begin{equation}\label{4d4}
\mu\frac{d}{d\mu}S_{EE} = 4((a_{UV} - a_{IR}) - \frac{c_{UV} - c_{IR}}{3}) + ..........
\end{equation}
We have used the results of \cite{Fursaev:1995ef} to simplify the integration of various geometric invariants on the cone. We have omitted the terms which come from the weyl-invariant part of the dilaton effective action. In \cite{Banerjee:2014daa} we discussed a class of such terms which give rise to interesting universal terms in entanglement entropy which were observed for example in \cite{Hertzberg:2010uv,Hertzberg:2012mn,Hung:2011ta,Lewkowycz:2012qr}. The term $(\ref{4d4})$ we have written down involving the central charges does not receive any contribution from the Weyl invariant part of the dilaton effective action as was shown in scetion-$8$ of \cite{Banerjee:2014daa}. \footnote{Actually this way of calculating the entanglement entropy is well suited for computing the renormalized entanglement entropy as defined in \cite{Liu:2012eea}. it will be very interesting to see if one could reproduce the various properties of this which were obtained from holographic analysis.}

$(\ref{4d4})$ can be thought of as the four dimensional generalisation of the Calabrese-Cardy \cite{Calabrese:2004eu} answer $(\ref{cc})$. This formula can be easily generalised to any even dimension using this method.

We can go back to the expression for the entanglement entropy, $S_{EE}$, starting from $(\ref{4d4})$,
\begin{equation}
S_{EE} = 4(a_{UV} - \frac{c_{UV}}{3}) ln(\mu a) - 4(a_{IR} - \frac{c_{IR}}{3}) ln(\mu R) + ............
\end{equation}
where $a$ is the short-distance cut-off.

%\section{Renormalized Entanglement Entropy} 
%In \cite{Liu:2012eea} the authors defined the renormalized entanglement entropy, which in even dimensions is given by,
%\begin{equation}
%S_{d}^{\Sigma}(R) = \frac{1}{(d-2)!!} (R\frac{d}{dR} - 2)........(R\frac{d}{dR} -(d-2)) R\frac{d}{dR} S^{\Sigma}(R)
%\end{equation}
%where $d$ is the total space-time dimensions. $\Sigma$ is the entangling surface and $R$ is a length scale associated with the surface.\footnote{The authors of \cite{Liu:2012eea} considered a scalable entangling surface. Please see the reference for details on the geometry of these types of surfaces.} The quantity, $S^{\Sigma}_d(R)$, has some nice features which were proved in \cite{Liu:2012eea}. For example this quantity is UV finite and for a renormalizable quantum field theory it interpolates between its value at the UV fixed point and the IR fixed point as $R$ is increased form zero to infinity. Moreover its value at the fixed point is independent of $R$. 

\section{Theories Which Flow Between Scale Invariant But Not Conformally Invariant Fixed Points}
In \cite{Banerjee:2014daa} we argued that the method computing the entanglement entropy using a constant dilaton background field is also valid if we replace the CFTs in the UV and the IR by only scale invariant theories (SFT). The only change is that the expression for the scale anomaly or the integrated trace of the energy momentum changes and in four dimensions one gets an extra term \cite{Dymarsky:2014zja,Nakayama:2011wq} which can be written as, 
\begin{equation}
e \int d^4x \sqrt{h} R^2(h) 
\end{equation} 
where $R(h)$ is the Ricci scalar constructed out of the background metric $h$ and $e$ is a pure number. This gives rise to an extra term in the universal part of the dilaton effective action, of the form,
\begin{equation}
F_{extra}(n,\tau) \sim \ \tau \ (e_{UV} - e_{IR}) \int_{cone} d^4x \sqrt{h} R^2(h)
\end{equation}
This extra tem gives a contribution to the entanglement entropy which is not there if the fixed points are CFTs rather than SFTs. \textit{It will be very interesting to see what property of entanglement entropy prohibits the presence of this term when the fixed points are CFTs. This may also point to a strategy for proving if every unitary scale invariant theory is conformally invariant or not, using the properties of entanglement entropy. This may turn out to be useful in say six dimensions where no such proof exists so far.} \footnote{For example the proof of F-theorem in three dimensions was done using the strong sub additivity property of entanglement entropy \cite{Casini:2012ei}.} \footnote{Please see \cite{Polchinski:1987dy,Dymarsky:2013pqa,Dymarsky:2014zja,Bzowski:2014qja} for recent developments on the subject.}

%\subsection{Massive scalar in four dimensions}
%Using the methods of squashed cones the reference \cite{Fursaev:2013fta} made a proposal for the logarithmically divergent terms in the entanglement entropy of a massive scalar field of mass $m$. 
\section{Discussion}
It will be very interesting to prove these results from a holographic calculation of entanglement entropy \cite{Ryu:2006bv,Lewkowycz:2013nqa} in RG flow geometries \cite{Freedman:1999gp,Khavaev:1998fb,Girardello:1999bd,Lin:2004nb}. The universal coefficients of the logarithmically divergent terms involving the UV central charges match with the answers obtained from holography \cite{Hung:2011ta}. It will be very interesting to reproduce the contribution which depends on IR central charges. Another interesting direction to pursue will be to generalise this technique to the calculation of entanglement entropy of excited states, for which interesting results were obtained recently \cite{Nozaki:2014hna, Caputa:2014vaa}.

\acknowledgments 
I am grateful to John Cardy for very helpful correspondence and a detailed explanation of the two dimensional result when the flow is towards a non-trivial IR fixed point and the problem of IR divergence in this case. I would also like to thank Tadashi Takayanagi for his insightful questions which led to a more careful analysis of this problem. This work was supported by World Premier International Research Center Initiative (WPI), MEXT, Japan.


\begin{thebibliography} {99}

%\bibitem{Srednicki:1993im}
 % M.~Srednicki,
 % ``Entropy and area,''
 % Phys.\ Rev.\ Lett.\  {\bf 71}, 666 (1993)
 % [arXiv:hep-th/9303048].
  
  \bibitem{Holzhey:1994we}
  C.~Holzhey, F.~Larsen and F.~Wilczek,
  ``Geometric and renormalized entropy in conformal field theory,''
  Nucl.\ Phys.\  B {\bf 424}, 443 (1994)
  [arXiv:hep-th/9403108].
 
 \bibitem{Calabrese:2004eu}
  P.~Calabrese and J.~L.~Cardy,
  ``Entanglement entropy and quantum field theory,''
  J.\ Stat.\ Mech.\  {\bf 0406}, P06002 (2004)
  [arXiv:hep-th/0405152].
      
  \bibitem{Calabrese:2009qy}
  P.~Calabrese and J.~Cardy,
  ``Entanglement entropy and conformal field theory,''
  J.\ Phys.\ A  {\bf 42}, 504005 (2009)
  [arXiv:0905.4013 [cond-mat.stat-mech]].
  
  \bibitem{Komargodski:2011vj} 
  Z.~Komargodski and A.~Schwimmer,
  ``On Renormalization Group Flows in Four Dimensions,''
  JHEP {\bf 1112}, 099 (2011)
  [arXiv:1107.3987 [hep-th]].
  
  \bibitem{Komargodski:2011xv} 
  Z.~Komargodski,
  ``The Constraints of Conformal Symmetry on RG Flows,''
  JHEP {\bf 1207}, 069 (2012)
  [arXiv:1112.4538 [hep-th]].
  
  \bibitem{Luty:2012ww} 
  M.~A.~Luty, J.~Polchinski and R.~Rattazzi,
  ``The $a$-theorem and the Asymptotics of 4D Quantum Field Theory,''
  JHEP {\bf 1301}, 152 (2013)
  [arXiv:1204.5221 [hep-th]].
  
  \bibitem{Elvang:2012st} 
  H.~Elvang, D.~Z.~Freedman, L.~-Y.~Hung, M.~Kiermaier, R.~C.~Myers and S.~Theisen,
  ``On renormalization group flows and the a-theorem in 6d,''
  JHEP {\bf 1210}, 011 (2012)
  [arXiv:1205.3994 [hep-th]].
  
  \bibitem{Schwimmer:2010za} 
  A.~Schwimmer and S.~Theisen,
  ``Spontaneous Breaking of Conformal Invariance and Trace Anomaly Matching,''
  Nucl.\ Phys.\ B {\bf 847}, 590 (2011)
  [arXiv:1011.0696 [hep-th]].
  
   \bibitem{Polchinski:1987dy} 
  J.~Polchinski,
  ``Scale and Conformal Invariance in Quantum Field Theory,''
  Nucl.\ Phys.\ B {\bf 303}, 226 (1988).
    
  \bibitem{Dymarsky:2013pqa} 
  A.~Dymarsky, Z.~Komargodski, A.~Schwimmer and S.~Theisen,
  ``On Scale and Conformal Invariance in Four Dimensions,''
  arXiv:1309.2921 [hep-th].
  
  \bibitem{Dymarsky:2014zja} 
  A.~Dymarsky, K.~Farnsworth, Z.~Komargodski, M.~A.~Luty and V.~Prilepina,
  ``Scale Invariance, Conformality, and Generalized Free Fields,''
  arXiv:1402.6322 [hep-th].
  
  \bibitem{Bzowski:2014qja} 
  A.~Bzowski and K.~Skenderis,
  ``Comments on scale and conformal invariance in four dimensions,''
  arXiv:1402.3208 [hep-th].
            
 \bibitem{Nakayama:2011wq} 
  Y.~Nakayama,
  ``On epsilon-conjecture in a-theorem,''
  Mod.\ Phys.\ Lett.\ A {\bf 27}, 1250029 (2012)
  [arXiv:1110.2586 [hep-th]].
  
\bibitem{Banerjee:2014daa} 
  S.~Banerjee,
  ``Trace Anomaly Matching and Exact Results For Entanglement Entropy,''
  arXiv:1405.4876 [hep-th].
  
  \bibitem{Casini:2009sr} 
  H.~Casini and M.~Huerta,
  ``Entanglement entropy in free quantum field theory,''
  J.\ Phys.\ A {\bf 42}, 504007 (2009)
  [arXiv:0905.2562 [hep-th]]
          
   \bibitem{Ryu:2006bv}
  S.~Ryu and T.~Takayanagi,
  ``Holographic derivation of entanglement entropy from AdS/CFT,''
  Phys.\ Rev.\ Lett.\  {\bf 96}, 181602 (2006)
  [arXiv:hep-th/0603001].
  
  \bibitem{Ryu:2006ef}
  S.~Ryu and T.~Takayanagi,
  ``Aspects of Holographic Entanglement Entropy,''
  JHEP {\bf 0608}, 045 (2006)
  [arXiv:hep-th/0605073].
  
 % \bibitem{Casini:2011kv}
 % H.~Casini, M.~Huerta and R.~C.~Myers,
 % ``Towards a derivation of holographic entanglement entropy,''
 % JHEP {\bf 1105}, 036 (2011)
 % [arXiv:1102.0440 [hep-th]]. 
  
  \bibitem{Lewkowycz:2013nqa} 
  A.~Lewkowycz and J.~Maldacena,
  ``Generalized gravitational entropy,''
  JHEP {\bf 1308}, 090 (2013)
  [arXiv:1304.4926 [hep-th]].
  
 % \bibitem{Hartman:2013mia} 
  %T.~Hartman,
  %``Entanglement Entropy at Large Central Charge,''
  %arXiv:1303.6955 [hep-th].
  
  %\bibitem{Faulkner:2013yia} 
  %T.~Faulkner,
  %``The Entanglement Renyi Entropies of Disjoint Intervals in AdS/CFT,''
  %arXiv:1303.7221 [hep-th].
  
 % \bibitem{Faulkner:2013ana} 
 % T.~Faulkner, A.~Lewkowycz and J.~Maldacena,
%  ``Quantum corrections to holographic entanglement entropy,''
  %JHEP {\bf 1311}, 074 (2013)
 % [arXiv:1307.2892].
  
 %  \bibitem{Solodukhin:2008dh}
%  S.~N.~Solodukhin,
 % ``Entanglement entropy, conformal invariance and extrinsic geometry,''
 % Phys.\ Lett.\  B {\bf 665}, 305 (2008)
%  [arXiv:0802.3117 [hep-th]].
    
%  \bibitem{Barrella:2013wja} 
 % T.~Barrella, X.~Dong, S.~A.~Hartnoll and V.~L.~Martin,
 % ``Holographic entanglement beyond classical gravity,''
 % JHEP {\bf 1309}, 109 (2013)
  %[arXiv:1306.4682 [hep-th]].
  
 % \bibitem{Cardy:2014jwa} 
 % J.~Cardy and C.~P.~Herzog,
%  ``Universal Thermal Corrections to Single Interval Entanglement Entropy for Conformal Field Theories,''
 % Phys.\ Rev.\ Lett.\  {\bf 112}, 171603 (2014)
 % [arXiv:1403.0578 [hep-th]].
  
  %\bibitem{Lewkowycz:2013laa} 
  %A.~Lewkowycz and J.~Maldacena,
  %``Exact results for the entanglement entropy and the energy radiated by a quark,''
  %JHEP {\bf 1405}, 025 (2014)
  %[arXiv:1312.5682 [hep-th]].
  
  %\bibitem{Datta:2013hba} 
  %S.~Datta and J.~R.~David,
  %``RŽnyi entropies of free bosons on the torus and holography,''
  %JHEP {\bf 1404}, 081 (2014)
  %[arXiv:1311.1218 [hep-th]].
  
  %\bibitem{Perlmutter:2013gua} 
  %E.~Perlmutter,
  %``A universal feature of CFT RŽnyi entropy,''
  %JHEP {\bf 1403}, 117 (2014)
  %[arXiv:1308.1083 [hep-th]].
  
%  \bibitem{Rosenhaus:2014woa} 
 % V.~Rosenhaus and M.~Smolkin,
  %``Entanglement Entropy: A Perturbative Calculation,''
  %arXiv:1403.3733 [hep-th].
      
  \bibitem{Myers:2010tj}
  R.~C.~Myers and A.~Sinha,
  ``Holographic c-theorems in arbitrary dimensions,''
  JHEP {\bf 1101}, 125 (2011)
  [arXiv:1011.5819 [hep-th]].
  
 % \bibitem{Myers:2010xs} 
 % R.~C.~Myers and A.~Sinha,
  %``Seeing a c-theorem with holography,''
 % Phys.\ Rev.\ D {\bf 82}, 046006 (2010)
  %[arXiv:1006.1263 [hep-th]].
  
\bibitem{Casini:2012ei} 
  H.~Casini and M.~Huerta,
  ``On the RG running of the entanglement entropy of a circle,''
  Phys.\ Rev.\ D {\bf 85}, 125016 (2012)
  [arXiv:1202.5650 [hep-th]].
  
    \bibitem{Liu:2012eea} 
  H.~Liu and M.~Mezei,
  ``A Refinement of entanglement entropy and the number of degrees of freedom,''
  JHEP {\bf 1304}, 162 (2013)
  [arXiv:1202.2070 [hep-th]].
  
\bibitem{Hertzberg:2010uv} 
  M.~P.~Hertzberg and F.~Wilczek,
  ``Some Calculable Contributions to Entanglement Entropy,''
  Phys.\ Rev.\ Lett.\  {\bf 106}, 050404 (2011)
  [arXiv:1007.0993 [hep-th]].
  
  \bibitem{Hertzberg:2012mn} 
  M.~P.~Hertzberg,
  ``Entanglement Entropy in Scalar Field Theory,''
  J.\ Phys.\ A {\bf 46}, 015402 (2013)
  [arXiv:1209.4646 [hep-th]]. 
  
 \bibitem{Hung:2011ta}
  L.~Y.~Hung, R.~C.~Myers and M.~Smolkin,
  ``Some Calculable Contributions to Holographic Entanglement Entropy,''
  arXiv:1105.6055 [hep-th].  
  
  \bibitem{Lewkowycz:2012qr} 
  A.~Lewkowycz, R.~C.~Myers and M.~Smolkin,
  ``Observations on entanglement entropy in massive QFT's,''
  JHEP {\bf 1304}, 017 (2013)
  [arXiv:1210.6858 [hep-th]].
  
  \bibitem{Freedman:1999gp} 
  D.~Z.~Freedman, S.~S.~Gubser, K.~Pilch and N.~P.~Warner,
  ``Renormalization group flows from holography supersymmetry and a c theorem,''
  Adv.\ Theor.\ Math.\ Phys.\  {\bf 3}, 363 (1999)
  [hep-th/9904017].  
  
  \bibitem{Khavaev:1998fb} 
  A.~Khavaev, K.~Pilch and N.~P.~Warner,
  ``New vacua of gauged N=8 supergravity in five-dimensions,''
  Phys.\ Lett.\ B {\bf 487}, 14 (2000)
  [hep-th/9812035].
  
 \bibitem{Girardello:1999bd} 
  L.~Girardello, M.~Petrini, M.~Porrati and A.~Zaffaroni,
  ``The Supergravity dual of N=1 superYang-Mills theory,''
  Nucl.\ Phys.\ B {\bf 569}, 451 (2000)
  [hep-th/9909047]. 
  
 \bibitem{Lin:2004nb} 
  H.~Lin, O.~Lunin and J.~M.~Maldacena,
  ``Bubbling AdS space and 1/2 BPS geometries,''
  JHEP {\bf 0410}, 025 (2004)
  [hep-th/0409174]. 
  
  \bibitem{Nozaki:2014hna} 
  M.~Nozaki, T.~Numasawa and T.~Takayanagi,
  ``Quantum Entanglement of Local Operators in Conformal Field Theories,''
  Phys.\ Rev.\ Lett.\  {\bf 112}, 111602 (2014)
  [arXiv:1401.0539 [hep-th]].
  
    
  \bibitem{Caputa:2014vaa} 
  P.~Caputa, M.~Nozaki and T.~Takayanagi,
  ``Entanglement of Local Operators in large N CFTs,''
  arXiv:1405.5946 [hep-th].
  
  
 % \bibitem{Bhattacharyya:2012tc} 
 % A.~Bhattacharyya, L.~-Y.~Hung, K.~Sen and A.~Sinha,
  %``On c-theorems in arbitrary dimensions,''
 % Phys.\ Rev.\ D {\bf 86}, 106006 (2012)
  %[arXiv:1207.2333 [hep-th]].
  
 % \bibitem{Castro:2014tta} 
 % A.~Castro, S.~Detournay, N.~Iqbal and E.~Perlmutter,
  %``Holographic entanglement entropy and gravitational anomalies,''
  %arXiv:1405.2792 [hep-th].  
  
  %\bibitem{Jafferis:2010un} 
  %D.~L.~Jafferis,
  %``The Exact Superconformal R-Symmetry Extremizes Z,''
  %JHEP {\bf 1205}, 159 (2012)
  %[arXiv:1012.3210 [hep-th]]
  
% \bibitem{Jafferis:2011zi} 
%  D.~L.~Jafferis, I.~R.~Klebanov, S.~S.~Pufu and B.~R.~Safdi,
 % ``Towards the F-Theorem: N=2 Field Theories on the Three-Sphere,''
 % JHEP {\bf 1106}, 102 (2011)
  %[arXiv:1103.1181 [hep-th]].
  
 
% \bibitem{Klebanov:2012va} 
 % I.~R.~Klebanov, T.~Nishioka, S.~S.~Pufu and B.~R.~Safdi,
 % ``Is Renormalized Entanglement Entropy Stationary at RG Fixed Points?,''
 % JHEP {\bf 1210}, 058 (2012)
  %[arXiv:1207.3360 [hep-th]].
  
 % \bibitem{Klebanov:2011td} 
 % I.~R.~Klebanov, S.~S.~Pufu, S.~Sachdev and B.~R.~Safdi,
 % ``Entanglement Entropy of 3-d Conformal Gauge Theories with Many Flavors,''
 % JHEP {\bf 1205}, 036 (2012)
  %[arXiv:1112.5342 [hep-th]]. 
  
 % \bibitem{Agon:2013iva} 
 % C.~A.~Agon, M.~Headrick, D.~L.~Jafferis and S.~Kasko,
  %``Disk entanglement entropy for a Maxwell field,''
 % Phys.\ Rev.\ D {\bf 89}, 025018 (2014)
  %[arXiv:1310.4886 [hep-th]].
      
%  \bibitem{Hung:2011xb}
%  L.~Y.~Hung, R.~C.~Myers and M.~Smolkin,
%  ``On Holographic Entanglement Entropy and Higher Curvature Gravity,''
 % JHEP {\bf 1104}, 025 (2011)
 % [arXiv:1101.5813 [hep-th]]. 
  
%  \bibitem{deBoer:2011wk}
%  J.~de Boer, M.~Kulaxizi and A.~Parnachev,
 % ``Holographic Entanglement Entropy in Lovelock Gravities,''
 % arXiv:1101.5781 [hep-th].
      
  \bibitem{Fursaev:1995ef} 
  D.~V.~Fursaev and S.~N.~Solodukhin,
  ``On the description of the Riemannian geometry in the presence of conical defects,''
  Phys.\ Rev.\ D {\bf 52}, 2133 (1995)
  [hep-th/9501127].
  
  %\bibitem{Fursaev:2013fta} 
 % D.~V.~Fursaev, A.~Patrushev and S.~N.~Solodukhin,
  %``Distributional Geometry of Squashed Cones,''
 % Phys.\ Rev.\ D {\bf 88}, no. 4, 044054 (2013)
 % [arXiv:1306.4000 [hep-th]].
  
 
  
%  \bibitem{Solodukhin:2013yha} 
 % S.~N.~Solodukhin,
 % ``The a-theorem and entanglement entropy,''
 % arXiv:1304.4411 [hep-th].  
  
   \end{thebibliography}
\end{document}